\documentstyle[prl,aps,multicol,epsf,amssymb]{revtex}

\begin{document}

\draft

\title {The relative importance of electron-electron interactions compared to disorder in the two-dimensional ``metallic'' state}

\author {A.~Lewalle ,$^{1}$ M.~Pepper,$^{1}$ C.~J.~B.~Ford,$^{1}$ E.~H.~Hwang,$^2$ S.~Das~Sarma,$^2$ D.~J.~Paul,$^{1}$ and G. Redmond$^{3}$}

\address {$^{1}$ Cavendish Laboratory, Madingley Road, Cambridge CB3 0HE, United Kingdom}
\address {$^2$ Department of Physics, University of Maryland, College Park, Maryland 20742-4111}
\address {$^{3}$ Nanotechnology Group, National Microelectronics Research Centre, Lee Maltings, Prospect Row, Cork, Ireland}

\date {\today}
\maketitle
\widetext

\begin {abstract}   
\leftskip 54.8pt
\rightskip 54.8pt

The effect of substrate bias and surface gate voltage on the low temperature resistivity of a Si-MOSFET is studied for electron concentrations where the resistivity increases with increasing temperature.  This technique offers two degrees of freedom for controlling the electron concentration and the device mobility, thereby providing a means to evaluate the relative importance of electron-electron interactions and disorder in this so-called  ``metallic'' regime. For temperatures well below the Fermi temperature, the data obey a scaling law where the disorder parameter ($k_{\rm{F}}l$), and not the concentration, appears explicitly.  This suggests that interactions, although present, do not alter the Fermi-liquid properties of the system fundamentally. Furthermore, this experimental observation is reproduced in results of calculations based on temperature-dependent screening,  in the context of Drude-Boltzmann theory.

\end {abstract}

\begin {multicols} {2}

\narrowtext
The term ``metallic behaviour'', for a two-dimensional (2D) system with concentration $n$ greater than some critical value $n_{\rm{c}}$, has come  to designate a drop in the resistivity $\rho$ as the temperature $T$ is decreased, as opposed to ``insulating behaviour'' (for $n<n_{\rm{c}}$) for which $\rho$ increases exponentially with decreasing temperature. In experiments performed in systems with low disorder and small values of $n_{\rm{c}}$ \cite{kravchenko1,kravchenko2,popovic,hanein,simmons,hamilton0,pudalov2,coleridge,papadakis,sarachik}, the change in $\rho$ in the metallic regime is considerably greater than was seen in earlier work at higher densities and in more disordered systems \cite{kawaguchi,hems}. There is still no consensus
on whether this change in $\rho$ is caused by Coulomb interactions or disorder.  The low values of $n_{\rm{c}}$ ($\propto r_{\rm{s}}^{-2}$, where $r_{\rm{s}}$ is the ratio of the Coulomb interaction to Fermi energy), the scaling behaviour of transport properties around $n_{\rm{c}}$ and the apparent saturation of $\rho (T)$ (with no sign of an upturn), seen by some authors \cite{kravchenko2,simmons,coleridge,sarachik,pudalov,kravchenko3,kravchenko4} as $T~\to~0$, have led to suggestions that the system is a true metal (in the sense that the charge carriers are delocalized at $T$~=~0).  The suggestion is that this novel phase results from strong electronic correlations, and that $n_{\rm{c}}$ marks a quantum critical point and a metal-insulator transition (MIT).  This contradicts the well-established principle that a 2D system of non-interacting carriers is insulating at $T$~=~0 \cite{abrahams}, but there is no definitive corresponding prediction when interactions are present.  Other authors, however, observe a negative magnetoresistance at low magnetic fields $B$ and weak-localization corrections to the resistivity, which are consistent with Fermi-liquid behaviour \cite{hamilton,simmons2,senz,uren}.  Many hypotheses have been put forward to explain the metallic behaviour, some invoking exotic interaction effects \cite{chakravarty,zhang,yoon} and others advocating a more traditional framework, suggesting that the ``metallic'' behaviour is only a finite-temperature effect that is overwhelmed by localization at sufficiently low temperatures \cite{altshuler,he,dassarma,dassarma2,klapwijk,meir}.  To a large extent, the debate hinges on the question of whether the relatively large values of $r_{\rm{s}}$ (typically $>10$), corresponding to $n_{\rm{c}}$, warrant fundamentally new physics.

In motivating our work, we note that as $n$ ($r_s$) decreases (increases),
it is not only the electron-electron interaction that is increasing in the 2D 
system, but also the effective disorder felt by the carriers. 
This is because the dominant disorder in semiconductor 2D systems arises 
from Coulomb scattering by charged impurities, which increases monotonically 
with decreasing density as the 2D carrier system becomes less effective in 
screening the Coulomb interaction between the carriers and the charged 
impurities. Since, in the absence of umklapp processes (not applicable 
in these systems), electron-electron interactions typically do not affect 
ohmic transport, there is good reason to believe that much of the observed 
``metallic'' behavior may be arising from the weakening of screening in 
the disorder potential, rather than from the increasing electron-electron 
interaction. By carefully studying temperature-dependent 2D transport in 
high quality Si-MOSFETs, where disorder strength and carrier density are 
controlled independently using a substrate bias, we hope to shed light 
in this Letter on this question of the relative importance of 
electron-electron interactions and the disorder potential in low 
density ``metallic'' transport.

A negative substrate bias ($V_{\rm{sub}}$) steepens the triangular confining potential normal to the surface, pushing the electron wave-function closer to the Si-SiO$_2$ interface.  This can either increase or decrease the mobility \cite{popovic,feng,hamilton2}.  The surface-gate voltage is used to tune $n$. Without assuming either a specific mechanism for the increase in $\rho$ with temperature or the existence of a quantum phase transition at $n_{\rm{c}}$, this technique probes the density-mobility phase space in a way that is impossible with singly-gated devices, and it allows a study of the metal in terms of $r_{\rm{s}}$ and the disorder parameter $k_{\rm{F}}l$, where $k_{\rm{F}}$ is the Fermi wave-vector and $l$ is the momentum relaxation length.  The value of $k_{\rm{F}}l$ can be calculated directly from the relation $\rho \equiv (h/2e^2)/(k_{\rm{F}}l)$, whilst $n$ is measured from the Hall effect at magnetic fields $B < 0.5~T$.

\begin{figure}
\epsfxsize=70truemm
\centerline{\epsffile{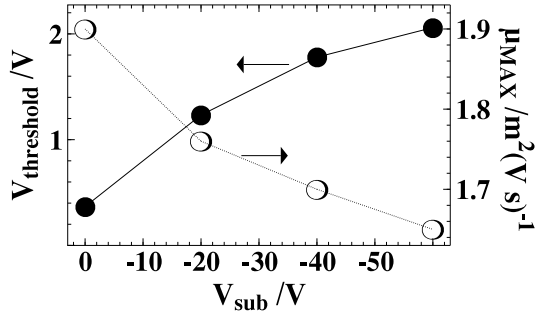}}
\vspace{6pt}
\caption {Threshold voltage and  peak mobility versus applied substrate bias for T=1.4~K.}
\label{fig:Fig1}
\end{figure}

Measurements were performed on an n-type Si-MOSFET inversion layer with a 200~nm-thick oxide layer, fabricated by NMRC, with a peak mobility (at $T$ = 1.4~K and with $V_{\rm{sub}}=0$) of 1.9~m$^2$V$^{-1}$s$^{-1}$ corresponding to $n = 4\times10^{15}~{\rm{ m}}^{-2}$ on $100~\Omega \rm{cm}$ substrate.  Devices were Hall bars of dimension $1000 \times 100~\rm{\mu m}^2$.  A $^4$He cryostat was used, covering a range $1.4~\rm{K} < T < 70~\rm{K}$.  Resistivities were measured using a standard a.c.\/four-terminal technique, with a constant source-drain current of $100~\rm{nA_{rms}}$ and a frequency of 19~Hz, after ensuring that the magnitude of the current did not induce significant electron heating above the substrate temperature and that the contacts were ohmic.  A substrate bias $V_{\rm{sub}}$ was applied at room temperature before the sample was lowered into the cryostat and cooled slowly.

One consequence of the (negative) substrate bias is to increase the threshold voltage, which results in the increase of the confining electric field, for a given $n$.  The peak mobility (to be  considered here as a rough measure of the  disorder) decreases as $V_{\rm{sub}}$ is made more negative, as shown in Fig.~\ref{fig:Fig1}, consistent with the enhancement of scattering from interface roughness and the charged impurities located near the SiO$_2$ interface. The enhancement of spin-orbit scattering is another possible consequence.  Negative magnetoresistance is observed in a weak magnetic field perpendicular to the system, however, in agreement with the prediction for the quenching of weak localization. No sign of the \it positive \rm magnetoresistance associated  with spin-orbit scattering is seen \cite{hikami}, indicating that this is not a strong effect. 
\begin{figure}
\epsfxsize=65truemm
\centerline{\epsffile{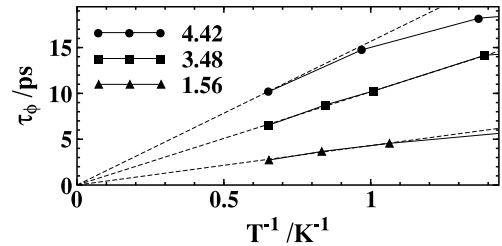}}
\vspace{6pt}
\caption {Phase decoherence time versus $T^{-1}$ for a concentrations 1.56, 3.48, and $4.42\times10^{15}~{\rm m}^{-2}$ at $V_{\rm sub} = 0$.}
\label{fig:Fig6}
\end{figure} 

The magnetoresistance demonstrates that weak localization, which results from quantum interference, is present in spite of the metallic behavior. Fig.~\ref{fig:Fig6} shows that the phase decoherence time $\tau_\phi$, obtained by fitting the negative magnetoresistance, is proportional to $T^{-1}$, in agreement with Fermi liquid  theory \cite{hikami,uren2}.  At sufficiently low temperatures, it is likely  that this localization will dominate.
\begin{figure}
\epsfxsize=85truemm
\centerline{\epsffile{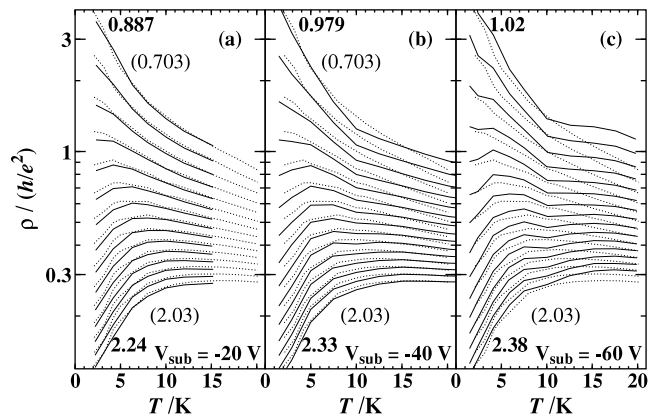}}
\vspace{6pt}
\caption {$\rho(T)$ over a range of densities for three substrate biases $-20$, $-40$, and $-60$ V (solid curves). Dotted curves show $\rho(T)$ for  $V_{\rm{sub}}~=~0$. Ranges of densities are  indicated  by bold numbers ($\times10^{15} ~\rm{m}^{-2}$) for $V_{\rm{sub}}~<~0$. Numbers in parentheses give the density range of the $V_{\rm{sub}}~=~0$ curves. The critical concentration $n_{\rm{c}}$ ranges from $0.99\times10^{15}~{\rm{ m}}^{-2}$ (for $V_{\rm{sub}}~=~0$) to $1.41\times10^{15}~{\rm{ m}}^{-2}$ (for $V_{\rm{sub}}~=~-60~\rm{V}$). Fermi temperature $T_{\rm F}$ [K] = 7.25 $n$ [$10^{15}~{\rm m}^{-2}$].}
\label{fig:Fig2}
\end{figure} 
The solid curves in Fig.~\ref{fig:Fig2} show $\rho (T)$ for different values of $V_{\rm{sub}}$ and for a range of densities spanning the ``metal''-insulator transition.  They are superposed on dotted curves, which correspond to $V_{\rm{sub}}$ = 0.  In each figure, the transition from strong localization ($d\rho/dT < 0$) to metallic behavior occurs at the same value of the resistivity (or $k_{\rm{F}}l$), $\rho_{\rm{c}} \approx h/e^2$, whereas $n_{\rm{c}}$ varies from $0.99\times10^{15}~\rm{m}^{-2}$ (at $V_{\rm{sub}}$ = 0) to $1.31\times10^{15}~\rm{m}^{-2}$ (at $V_{\rm{sub}}=-60~\rm{V}$).  This critical resistivity corresponds to $k_{\rm{F}}l$~=~0.5, a value below which the Fermi wavelength is poorly  defined.  It is worth emphasizing that it is $k_{\rm{F}}l$ and not $r_{\rm{s}}$ which is the critical parameter.
For densities greater than those shown in Fig.~\ref{fig:Fig2} (\it i.e. \rm $n > 2\times10^{15}~\rm{m}^{-2}$), $\rho(T)$ is non-monotonic, bending downwards in an insulator-like fashion as $T$ is increased beyond $T_{\rm{F}}$, the Fermi  temperature.  It also shows signs of saturation as $T$ approaches zero, possibly due to collision broadening or the finite Dingle temperature (discussed further below).  The intermediate temperature range shows an approximately linear dependence.  The gradients of the curves with $V_{\rm{sub}}~<~0$ differ from those of the $V_{\rm{sub}}~=~0$ reference curves, showing that the substrate bias does  not  only amount to a relabelling of the $\rho(T)$ curves.

With the hope of finding some underlying universality, we follow the work of A. R. Hamilton on GaAs \cite{hamilton3}, and consider theoretical studies, done in the context of Drude-Boltzmann theory, which take into account the strong temperature dependence of screening \cite{dassarma,dassarma3,gold}.  These calculations predict metallic behaviour resulting from the anomaly in the polarizability function $\Pi(q)$ at a value of the scattering wave vector $q~=~2k_{\rm{F}}$ which dominates the mobility.  To lowest order\cite{gold}, 
\begin{equation} 
\rho(T; n) = \rho(T~=~0; n) [1 + C(n) T/T_{\rm{F}}] 
\label{eq:goldeq}
\end{equation} 
\noindent valid in a range $T/T_{\rm{F}} \ll 1 \ll T/T_{\rm{D}}$, where $T_{\rm{D}}$ is the effective Dingle temperature and  $C$  is a function of $n$ and of the dominant scattering mechanism.

\begin{figure}
\epsfxsize=80truemm
\centerline{\epsffile{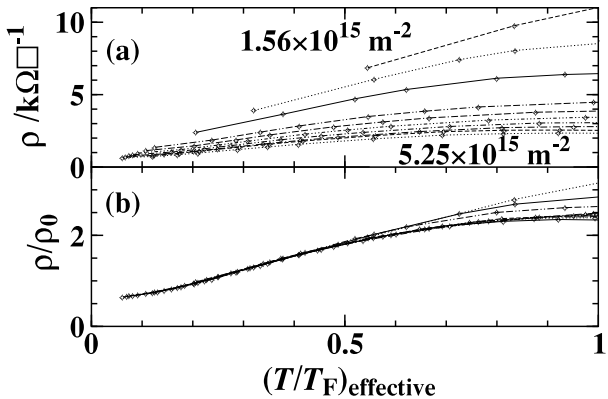}}
\vspace{6pt}
\caption {(a) Experimental $\rho(T/T_{\rm{F}})$ for a range of densities at a constant substrate bias, $V_{\rm{sub}}~=~0$. $T_{\rm{F}}$ ranges from 11.3~K to 38.0~K. The effective values of $T/T_{\rm{F}}$ displayed include the correction for collisional broadening (see Eq.~\ref{eq:correq}). (b) The same data with the vertical axis scaled by a factor $\rho_0$ determined  empirically for each curve.}
\label{fig:Fig3}
\end{figure} 

Following Eq.~\ref{eq:goldeq}, the temperature axis is normalized as $T/T_{\rm{F}}$, where  $T_{\rm{F}}$ is the Fermi temperature  calculated for each curve individually, as shown in Fig.~\ref{fig:Fig3}(a), for  $V_{\rm{sub}}~=~0$. The Dingle temperature was estimated from $T_{\rm D} = \hbar/2 \pi k_{\rm B} \tau_{\rm q}$, where $\tau_{\rm q}$ is the quantum lifetime, measured from the amplitude of Shubnikov-de Haas oscillations \cite{coleridge2}. For a range of concentrations $n=4.6-5.4\times10^{15}~\rm{m}^{-2}$ ($T_{\rm D}\sim35$~K), we obtain $T_{\rm D} \approx 1.1~\rm{K}$. At higher $T$, the curves are linear with a gradient that increases with decreasing $n$.  It is remarkable that another scaling factor $\rho_0$, applied to the \it vertical \rm axis and determined by eye for each curve, succeeds in collapsing the linear portion of all of the curves onto the bottom (high-density) curve (Fig.~\ref{fig:Fig3}b), at least for a range  of $T$ well away from $T_{\rm{F}}$.  Indeed, this scaling factor must equal the ratios of \it both \rm the gradients of the lines and their intercepts:
\begin{equation}
\rho(T; n)/\rho_0(n) = f(T/T_{\rm{F}}) ;\qquad T/T_{\rm{F}}~<~0.5.
\label{eq:scalingeq}
\end{equation}
This simple transformation holds well for $n>3\times10^{15}~\rm{m}^{-2}$ (for which $\rho (T = 1.4~\rm{K})~<~1.4~\rm{k\Omega}/\Box$) but fails at densities closer to the MIT, \it i.e. \rm it is impossible to match both the gradients and the intercepts with a single scaling factor.  A much better scaling of the curves is recovered by including a correction: 
\begin{equation} 
(T/T_{\rm{F}})_{\rm{effective}} = [(T/T_{\rm{F}})^2 + A/(k_{\rm{F}}l)^2]^{1/2} 
\label{eq:correq}. 
\end{equation} The value of  the coefficient $A$ was set to 1.0, but the overall quality of the fit was not sensitive to its exact value. The correction term $1/k_{\rm{F}}l$ may be interpreted as representing the broadening of the Fermi circle due to the finite scattering length (collisional broadening) \cite{dassarma3}.  Alternatively, it may be viewed as a lowering of the effective density (or $T_{\rm{F}}$) resulting from carrier freeze-out, a possibility envisaged by some authors \cite{dassarma2,klapwijk}.  With this correction, Eq.~\ref{eq:scalingeq} is obeyed down to $n = 1.56\times10^{15}~\rm{m}^{-2}$ (for which $\rho(T=1.4~\rm{K})=8~\rm{k\Omega/\Box}$), \it{i.e.}\rm~considerably closer to the MIT.  At even lower values of $n$, the correction term is much more significant compared with $T/T_{\rm{F}}$, so that the applicability of Eq.~\ref{eq:correq} becomes questionable.  Nevertheless, the success of Eq.~\ref{eq:correq} at slightly larger concentrations suggests that the low-$n$ regime close to the MIT is where the interaction between screening and disorder dominates and cannot be treated by a simple perturbative analysis.  For the higher-density data ($n \geq 3\times10^{15}~\rm{m}^{-2}$), the broadening correction to the temperature requires small modifications to the empirical $\rho_0$ values in order to achieve a satisfactory collapse, but the quality of the collapse is not significantly altered.
\begin{figure}
\epsfxsize=85truemm
\centerline{\epsffile{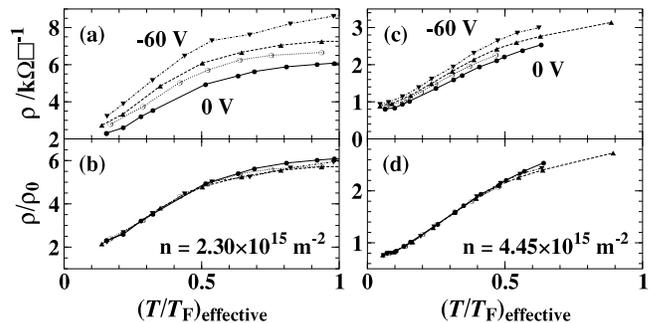}}
\vspace{6pt}
\caption {(a), (c) $\rho(T/T_{\rm{F}})$ for fixed densities $2.30\times10^{15}~\rm{m}^{-2}$ ($T_{\rm F}=16.7$~K) and $4.45\times10^{15}~\rm{m}^{-2}$ ($T_{\rm F}=32.3$~K) for substrate biases $0$, $-20$, $-40$, and  $-60$~V. (b) and (d) show the same data after scaling by $\rho_0(V_{\rm sub})$. }
\label{fig:Fig4}
\end{figure} 
\noindent Fig.~\ref{fig:Fig4} shows the same procedure, applied to a set of curves corresponding to different values of $V_{\rm{sub}}$ but with the same density (2.30 and $4.45\times 10^{15}~\rm{m}^{-2}$).  Again, Eq.~\ref{eq:scalingeq} is obeyed, but with $\rho_0=\rho_0(V_{\rm{sub}})$.
The similarity in the forms  of Eqs.~\ref{eq:goldeq} and \ref{eq:scalingeq} is notable,  but the scaling factors $\rho_0$ of Eq.~\ref{eq:scalingeq} may be equated with the $\rho(T=0)$ of Eq.~\ref{eq:goldeq} only in so far as (a) $\rho(T;n)$ remains linear outside the strict range $T\ll T_{\rm{F}}$, (b) the saturation observed at low temperatures may be discarded as due to the finite $T_{\rm{D}}$, and (c) $C$ is a sufficiently weak function of $n$. Experimentally, a constant $C \approx 8$ is observed, considerably greater than the expected theoretical values 2.0 - 3.0 \cite{gold}. In spite of this discrepancy, the experimental scaling behaviour suggests that the value of $\rho$ (or $k_{\rm{F}}l$) at finite $T$ is determined solely by its value at low $T$, with the concentration (and therefore $r_{\rm{s}}$) appearing only implicitly in $T_{\rm{F}}$. Combining Eq.~\ref{eq:scalingeq} with the identity $\rho \equiv (h/2e^2)/(k_{\rm{F}}l)$ we obtain
\begin{equation} 
\frac{k_{\rm{F}}l(T=0)}{k_{\rm{F}}l(T)} = \frac{f(T/T_{\rm{F}})}{f(0)}. 
\end{equation}

\begin{figure}
\epsfxsize=80truemm
\centerline{\epsffile{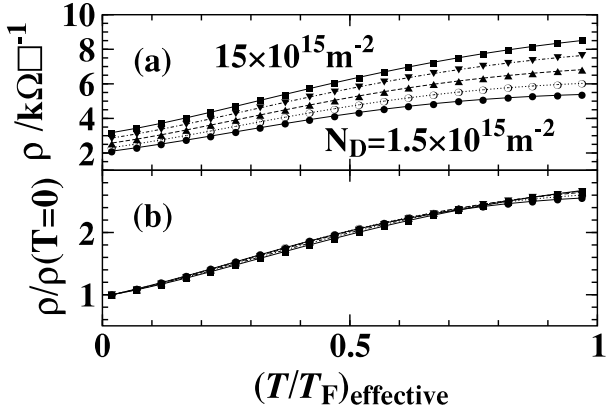}}
\vspace{6pt}
\caption {(a) Numerical calculation of $\rho(T/T_{\rm{F}})$ due to temperature-dependent screening, for constant $n = 2.3\times10^{15}~\rm{m}^{-2}$ ($T_{\rm{F}} = 16.7~\rm{K}$) and $N_{\rm{D}} = 1.5, 2.5, 4.5, 8.0$, and $15.0\times10^{15}~\rm{m}^{-2}$, with $T_{\rm{D}}/T_{\rm{F}} = 0, 0.06, 0.12, 0.18,$ and $0.24$ respectively. (b) The same data, with $\rho$ scaled by $\rho(T = 0)$.}
\label{fig:FigSDS}
\end{figure} 
Fig.~\ref{fig:FigSDS} shows numerical results of a numerical calculation of the temperature-dependent screening for experimental conditions analogousto those in Fig.~\ref{fig:Fig4}(a,b). It includes the temperature dependence of $\Pi(q)$ at finite temperature, the finite extent of the wavefunction in the direction normal to the interface (using a Fang-Howard distribution \cite{AFS}), and a variable effective disorder, parametrized by the the collision-induced Dingle temperature $T_{\rm{D}}$. (Here, $T_{\rm{D}}$ measures the strength of impurity  scattering and is, by definition, temperature-independent.) Carrier freeze-out \cite{dassarma2} was not assumed. The penetration depth of the wave function below the interface is determined by the depletion concentration, $N_{\rm{D}}$, which increases as $V_{\rm{sub}}$ is made more negative. The range of $N_{\rm{D}}$ indicated in Fig.~\ref{fig:FigSDS} is consistent with the range of $V_{\rm{sub}}$ ($0$ to $-60$~V) used in the experiment \cite{AFS}. By setting suitable values of $T_{\rm{D}}$, the calculation reproduces the experimental data with reasonable accuracy: (1) $\rho(T)$ increases in a metallic fashion, with a weaker temperature dependence appearing (at low $T$) as $T/T_{\rm{D}}$ decreases and (at high $T$) as $T/T_{\rm{F}}$ increases; (2) The scaling law, Eq.~\ref{eq:scalingeq}, is obeyed (with $\rho_0=\rho_0(N_{\rm{D}})$), although the ability to scale the theoretical curves as was done with the experimental data probably relies on the correct combination of $N_{\rm{D}}$ and $T_{\rm{D}}$. We emphasize, however, that the theoretical results presented here are not intended to give a fully quantitative description of the experimental results. Rather, we demonstrate that, to a large extent, temperature-dependent screening effects, in a conventional transport theory, 
can account for the main features of $\rho(T)$ in 
the so-called metallic regime in the absence of a magnetic field. 

In summary we have demonstrated that a simple emprical
scaling law, where the concentration does not appear  explicitly, is applicable on the metallic side of the MIT when either the concentration or the mobility is varied. (A similar effect has been observed in GaAs heterostructures \cite{hamilton3}.)  The quality of the scaling is improved, and is obeyed down to lower concentrations, by taking into account collision-broadening effects which enhance the effective temperature.  This suggests that electronic interactions are not the most crucial element in the metallic phenomenon, apart from screening the scattering potential, and that the ``metallic'' state is basically a Fermi liquid.  In fact, we observe a striking similarity with the prediction of Drude-Boltzmann transport theory in the presence of temperature-dependent screening (see also Ref. \cite{senz}).  Although quantitative agreement between this theory and our experimental results is only approximate, it is clear that many features of the ``metallic''  behavior can be understood in terms of screening and disorder, without resorting to electron-electron interaction or quantum interference effects.

This work was funded by the UK EPSRC.

\end{multicols}
\end{document}